\documentclass[a4paper,20pt]{article}
\usepackage[T1]{fontenc}
\usepackage{graphicx}
\usepackage{epsfig}
\usepackage{amsmath}
\usepackage{amsfonts}
\renewcommand{\abstract}{}
\begin{document}
\makeatletter
\renewcommand{\@oddhead}{\textit{YSC'17 Proceedings of Contributed Papers} \hfil 
\textit{A. Rozenkiewicz, K. Go\'zdziewski \& C. Migaszewski}}
\renewcommand{\@evenfoot}{\hfil \thepage \hfil}
\renewcommand{\@oddfoot}{\hfil \thepage \hfil}
\fontsize{11}{11} \selectfont
\newcommand\Chi{\sqrt{\chi^2_\nu}}
\title{\Large Modeling the Radial Velocities of HD~240210 with the Genetic Algorithms}
\author{\textsl{A. Rozenkiewicz$^{1}$, K. Go\'zdziewski$^{1}$ \& C. Migaszewski$^{1}$}}
\date{}
\maketitle
\begin{center} {\small $^{1}$Toru\'n Centre for Astronomy, 
Nicolaus Copernicus University, Gagarin Str. 11, 87-100 Toru\'n,  Poland\\
\{a.rozenkiewicz,k.gozdziewski,c.migaszewski\}@astri.umk.pl}
\end{center}
\begin{abstract}
More than 450 extrasolar planets are known to date.  To detect these intriguing
objects, many photometric and radial velocity (RV) surveys are in progress.  We
developed the Keplerian  FITting (KFIT) code, to model published and available
RV data. KFIT is based on a hybrid, quasi-global optimization technique relying
on the Genetic Algorithms and simplex algorithm. Here, we re-analyse the RV data
of evolved K3III star HD~240210. We confirm three equally good solutions which
might be interpreted as signals of 2-planet systems. Remarkably, one of these
best-fits describes  {\em long-term stable two-planet system}, involved in the
2:1~mean motion resonance (MMR).  It may be the first instance of this strong
MMR  in a multi-planet system hosted by evolved star, as the 2:1~MMR
configurations are already found around a few solar dwarfs.
\end{abstract}

\section*{Introduction}
Most of the discovered exoplanets are found around main sequence stars, perhaps
because the determination of stellar parameters is much easier than, for
instance for giant or active stars. The formation theory of planets hosted by
such stars is still developed and not understood well. Recently, there are only
known about of 10 exoplanets orbiting stars with masses greater than
2~$M_{\odot}$ (see, e.g., Extrasolar Planets Encyclopaedia,
http://exoplanet.eu). Some recent RV surveys focus on detecting planets around
the red giants, which are evolved main sequence stars. Unfortunately, the giants
and sub-giants are difficult targets for the RV technique. Usually, they are
chromospherically active, pulsating, surface-polluted by large spots, and
rotating slowly. The giants produce small number of sharp spectral lines and
their chromospheric activity and spots may change the profiles of spectral
lines. This intrinsic RV variability (also known as stellar ''jitter'') is
significantly larger than instrumental errors, and may be $\sim 20-30$~m/s and
larger. Actually, the stellar activity may even  mimic planetary signals (see,
e.g. \cite{berdyugina1995}). Hence, when interpreting the RV variability, we may
expect that different uncertainties and errors (usually, of unknown origin) may
significantly shift the best fits from the ``true'' solutions  in the parameters
space. In such a case, possibly global exploration of the parameter space and
the {\em dynamical stability} of multiple systems as an additional, {\em
implicit} observable may help us to correct and verify the derived best-fit
solutions for these factors,  and to conclude on the architecture of interacting
systems, even when limited data are available \cite{gozdziewski2008}. 

\section*{Keplerian model of the RV and optimization method}
We recall the {\em kinematic} RV model for the $N$-planet system, as the first
order approximation of the $N$-body model: 
\begin{equation}
V_{r}(t)=\sum_{i=1}^{N}K_{i}\left[\cos(\omega_{i} +\nu (t))+e_i \cos(\omega_{i} )\right]+V_{0},
\label{eq:K}
\end{equation}
where, for each planet in the system, $K_i$ is the semi-amplitude of the
signal,  $e_i$ is the orbital eccentricity,  $\omega_i$ is the argument of
pericenter,  $\nu_i(t)$ is the true anomaly, which depends on the orbital period
$P_i$, the time of periastron passage $\tau_i$ and eccentricity, and   $V_{0}$
is a constant instrumental offset.  The $N$-planet system is then characterized
by  $N_{p}= 5N + 1$ free parameters, to be determined from observations. Let us
note, that the inclinations and longitudes of nodes are not explicitly present
in Eq.~(\ref{eq:K}) and cannot be determined directly from the RV data alone, at
least in terms of the kinematic model in Eq.~(\ref{eq:K}).

To fit model Eq.~(\ref{eq:K}) to the RV data,  the Gaussian Least-Squares method
is  commonly used. To estimate  the best-fit model parameters, we seek for the
minimum of the reduced $\chi^2$ function, $\min \Chi({\bf p})$, which is
computed on the basis of synthetic signal $V_{r}(t_{i},{\bf p})$, where $t_i$
are moments of observations $V_{r,i}^{obs}$  with uncertainties $\sigma_{i}$
($i=1,\ldots,N_{RV}$ is the number of data), and  ${\bf p} \equiv (K_{p} , P_{p}
, e_{p} , \omega_{p} , \tau_{p} , V_{0})$, and $p = 1\equiv b, 2\equiv c,\ldots, N$
are for the model parameters of the $N$-planet system.

Clearly, even in the  kinematic formulation, $\Chi({\bf p})$ is a non-linear 
function. It is well known that it may exhibit numerous local extrema. Hence,
the exploration of multi-dimensional parameter space ${\bf p}$ requires
efficient, possibly global numerical optimization. In the past, we found that
good results in this problem may be achieved by an application of  the Genetic
Algorithms (GAs) \cite{charbonneau1995}, which mimic the biological evolution.
GAs search for the best fit solutions (best adapted population members) to the
model (to the environment) through ``breading'' an initial population
(parameters set), under particular genetic operators (e.g., cross-over,
mutation), and through a selection of gradually better adapted members. 
Although the GAs start with purely random initial ``population'', the search
converges to the best fit solutions \cite{charbonneau1995} by {\em
deterministic} way. Of course, the best fit solutions may be not unique. This is
very common in the case of modeling RV data. Overall, the GAs are robust and
quasi-global optimization technique, although they not provide efficiency and
accuracy of fast local methods, like the Levenberg-Marquardt algorithm. The GAs
are used in our hybrid KFIT code, developed for a few years now
\cite{gozdziewski2006},  which also makes use of the local, and accurate simplex
algorithm  \cite{nelder}. It helps to quickly refine  the final population of
the best fits, ``grown'' by the GAs. 

\section*{The results for the RV of HD~240210}
\begin{figure}
\centerline{
\vbox{
\hbox{
   \hbox{\includegraphics[width=0.5\textwidth]{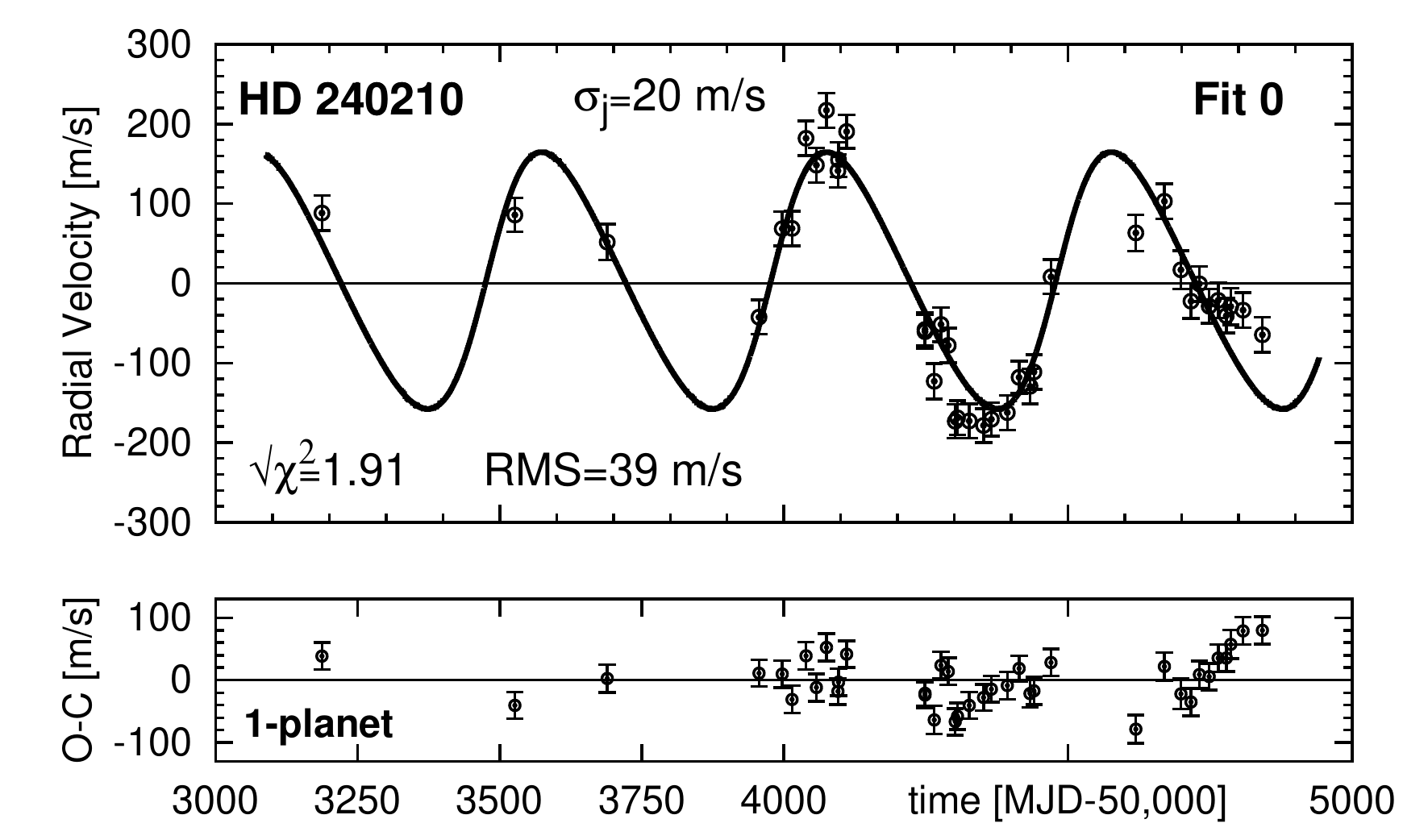}}
   \hbox{\includegraphics[width=0.5\textwidth]{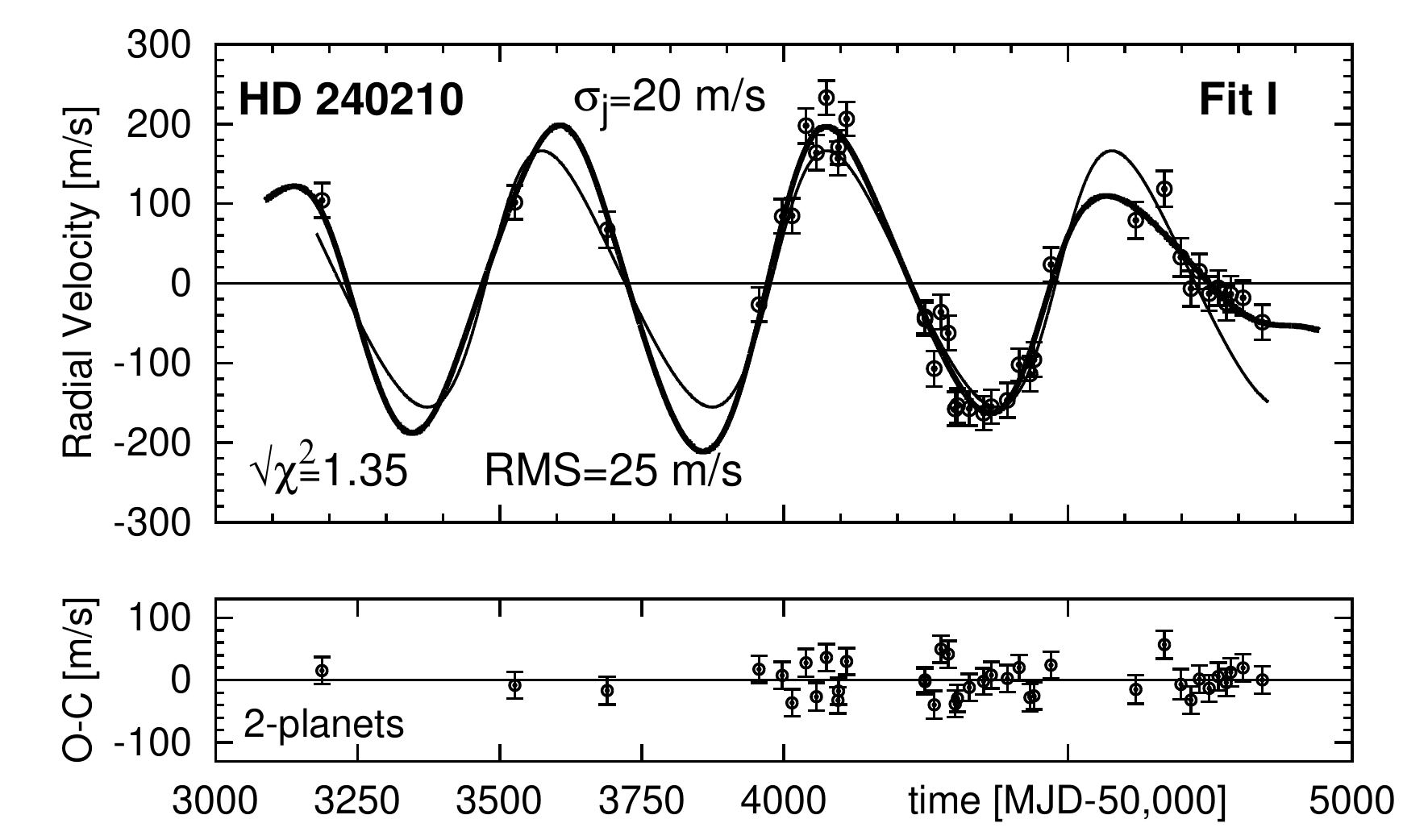}}
     }
\smallskip
\hbox{
   \hbox{\includegraphics[width=0.5\textwidth]{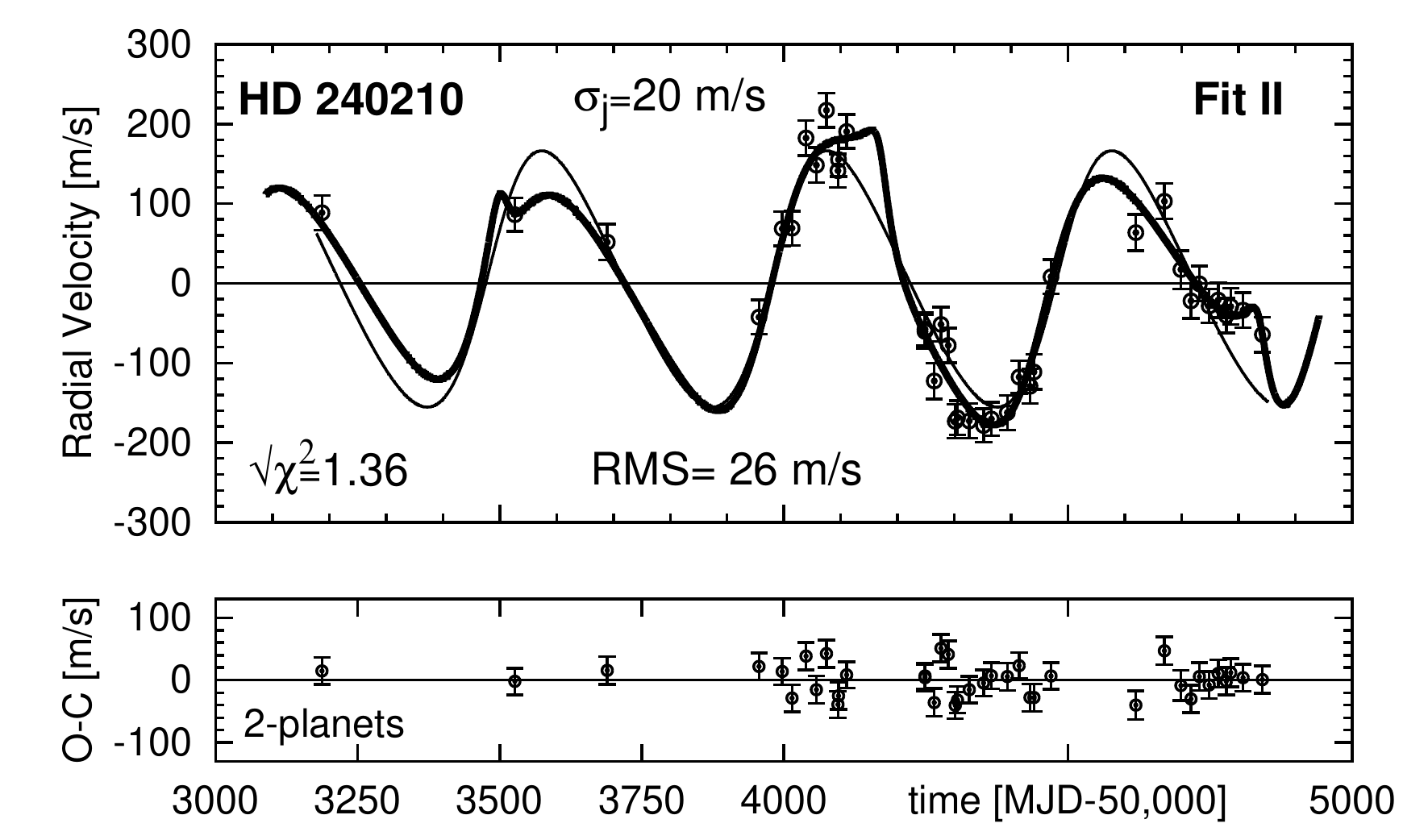}}
   \hbox{\includegraphics[width=0.5\textwidth]{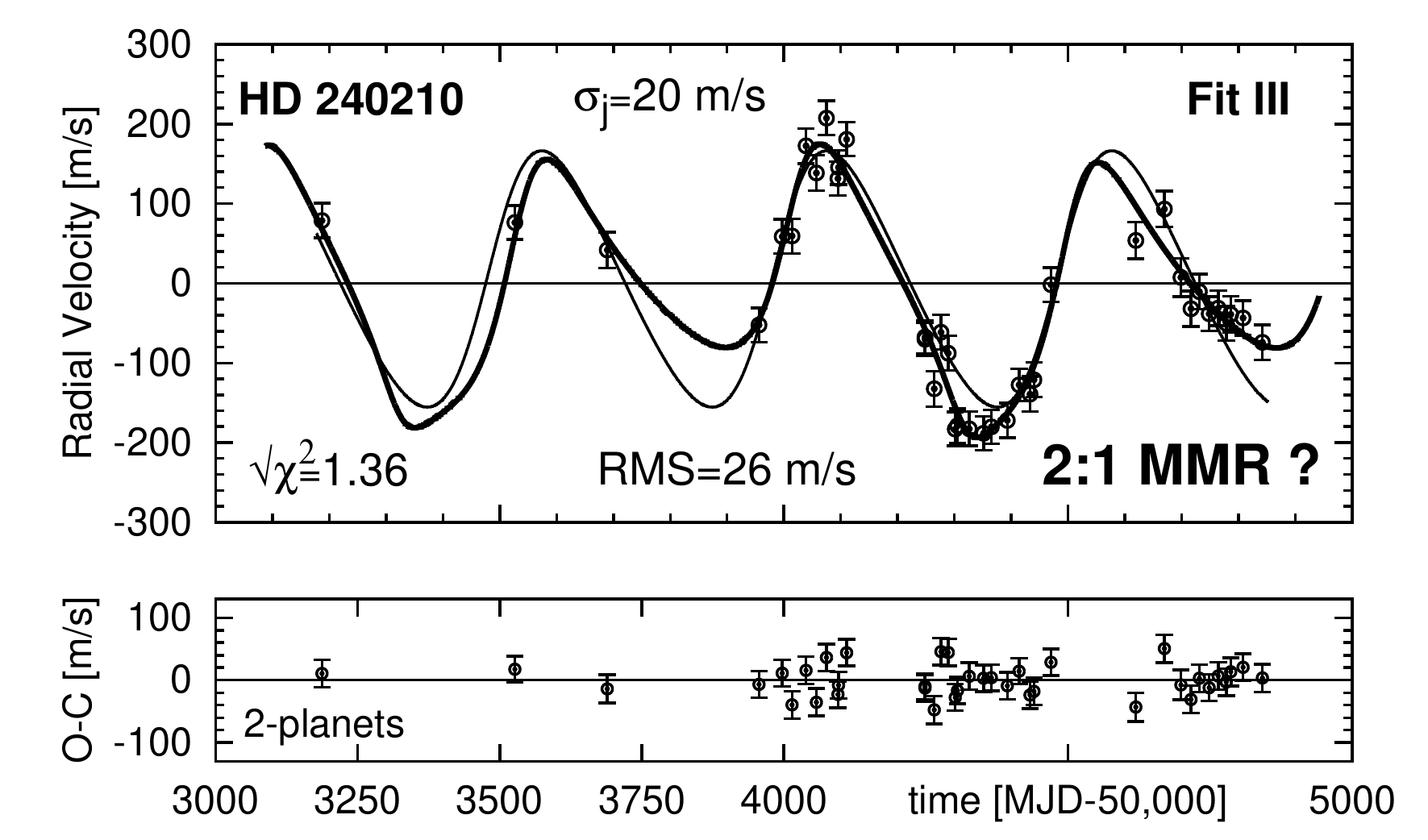}}
     }
}     
}
\caption{
The RV observations of HD~240210 \cite{niedzielski2010} and synthetic curves for
the best fit solutions found in this paper. {\em The left-upper panel}:  the
best 1-planet fit. Subsequent figures are for the best fits labeled with I, II,
III in Fig.~\ref{fig2}. For a reference, all these plots are accompanied by the
model curve of the 1-planet fit. See Table~1 for orbital elements.
}
\label{fig1}
\end{figure}

\begin{figure}
\centerline{
   \hbox{\includegraphics[width=0.66\textwidth]{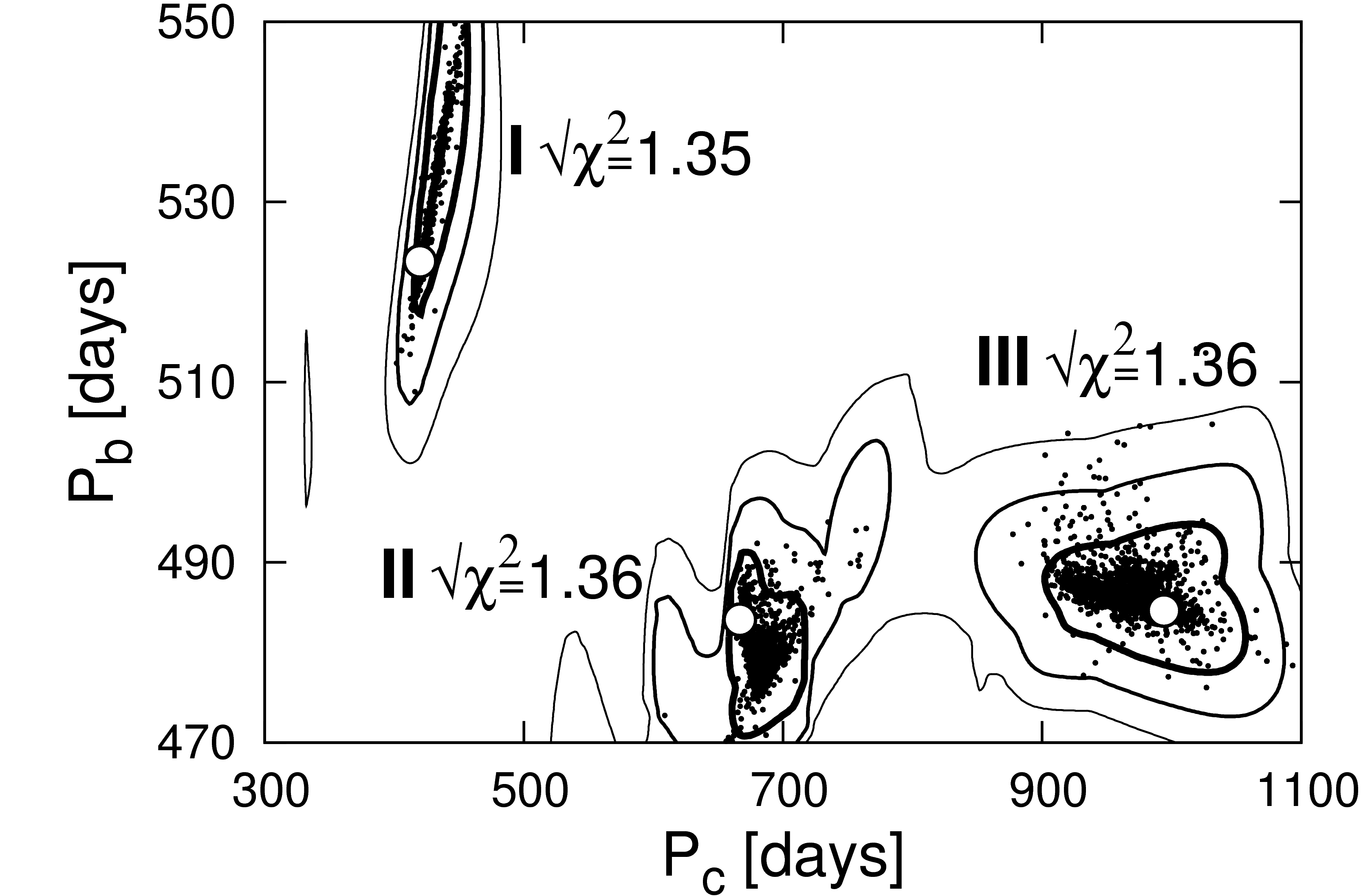}}
}
\caption{
Statistics of the Keplerian best fits gathered with the KFIT code,  projected
onto the ($P_b$, $P_c$)--plane of orbital periods.  Solutions marked with white
circles and labeled by I, II, III are for the best fits in different ``islands''
of the parameters space. Black filled circles are for all fits within formal
3$\sigma$-level of Fit~I; for a reference contours with 3,2,1 $\sigma$-level of
this solution, found through extensive, systematic scanning
\cite{gozdziewski2008} of the ($P_b$, $P_c$)-plane  with the Levenberg-Marquardt
algorithm, are plotted as curves of increasing thickness. See Table~1 for model
parameters.
}
\label{fig2}
\end{figure}

\begin{figure}
\centerline{
\hbox{
   \hbox{\includegraphics[width=0.46\textwidth]{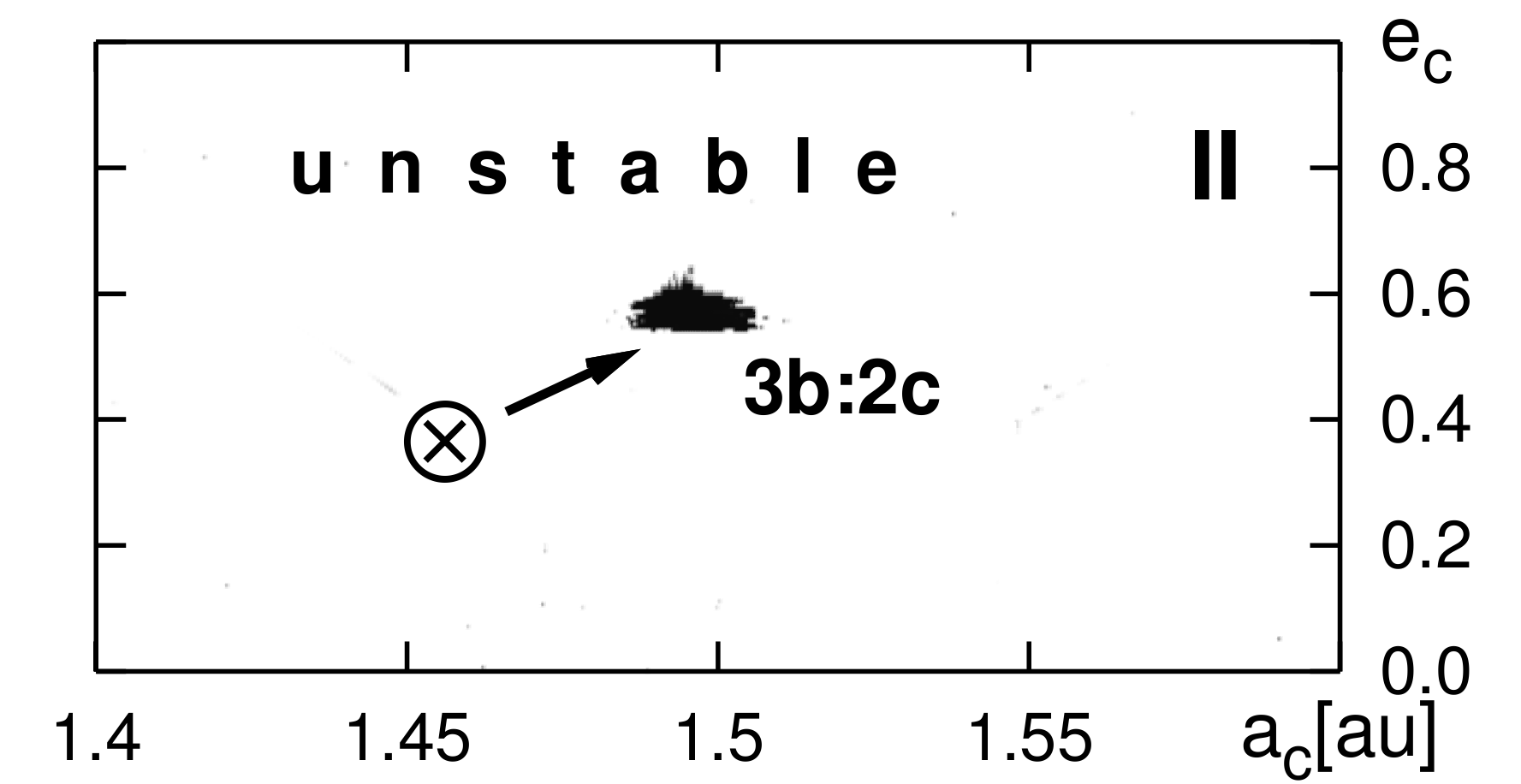}}
   \quad
   \hbox{\includegraphics[width=0.46\textwidth]{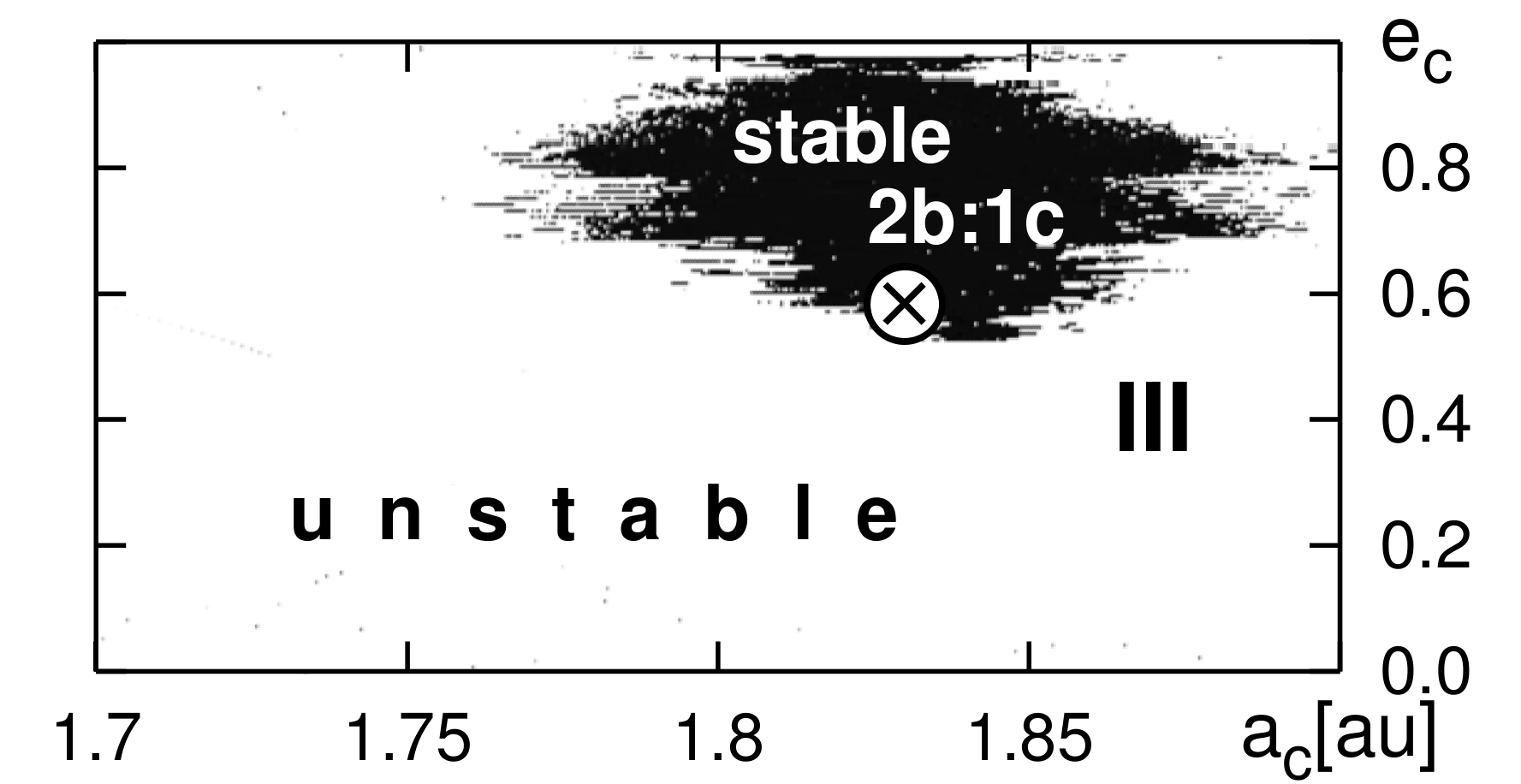}}
}
}
\caption{
Dynamical maps in terms of the MEGNO indicator \cite{gozdziewski2008} around
coplanar, edge-on, GAMP $N$-body fits (crossed circle): Fit~II  (3:2~MMR,
$\Chi=1.50$, rms $\sim 27$~m/s, {\em the left-hand  panel}), and  Fit III
(2:1~MMR, $\Chi=1.36$, rms $\sim 25$~m/s, {\em the right-hand panel}).  White
color is for unstable configurations, black is for stable solutions. Orbits  at
the epoch of the first RV of Fit~III in terms of 
$(m\,\mbox{[m}_{J}\mbox{]}, 
a\,\mbox{[au]}, e, \omega\,\mbox{[deg]}, {\cal M}\,\mbox{[deg]})$
are the following: 
(4.11,   1.14,   0.284, 304.8, 101.8)$_b$, 
(2.27,   1.83,   0.592, 162.3, 304.2)$_c$ for planets b, and c, respectively, 
$V_0=21.57$~m/s, 
stellar mass is 0.82~M$_{\odot}$ \cite{niedzielski2010};
Fit II is: 
(4.50,   1.143,   0.217, 249.24, 163.87)$_b$, 
(1.94,   1.506,   0.562, 53.09, 243.39)$_c$, 
$V_0=8.98$~m/s.
}
\label{fig3}
\end{figure}

In a recent work, Niedzielski~{\em et al.} \cite{niedzielski2010} detected a
planet hosted by evolved dwarf HD~240210. They found an excess in the rms of
Keplerian 1-planet model, and interpret this as possible signal of an additional
planet. We reproduced their 1-planet solution,  see Fig.~\ref{fig1} ({\em the
top-left panel}). The rms has significant scatter of $39$~m/s, compared to the
mean instrumental accuracy of $\sim 8$~m/s. This rms excess is difficult to
explain by the internal variability of the star, hence we tried to find a better
2-planet model. At first, we reproduced (see Fig.~\ref{fig1}, {\em the
top-right panel}) a tentative 2-planet solution  illustrated in the discovery paper.
However, we notice that the orbital periods are very similar $\sim 440$ and
$\sim 530$~days, respectively, indicating strong 4:3~MMR or 1:1~MMR. In such a
case, the kinematic model is not proper anymore, due to significant mutual
planetary  interactions. Moreover, such configurations could be hardly explained
by the planetary formation theory.
Still, having in mind that this solution is not unique, we performed an
extensive search for the local minima of $\Chi$ with the KFIT code. The
statistics of gathered fits is shown in Fig.~\ref{fig2}, through their
projection   onto orbital periods plane. In the range of orbital periods $\in
[60,3600]$~days, we found {\em three}, equally good best-fit models, which
correspond to different orbital configurations, and may be resolved at the
$2\sigma$ confidence level. In the parameter maps, the mentioned 2-planet model
is labeled as Fit~I. Two additional fits with orbital periods ratio close to 3:2
and 2:1 are labeled as Fit~II and Fit~III, respectively (see Table~1). Their
synthetic RV curves, with the RV of 1-planet model (thin curve) and observations
overplotted, are shown in subsequent panels of Fig.~\ref{fig1}. Note, that the
alternative 2-planet fits reduce the rms significantly, to $\sim 25$~m/s (i.e.,
by 1/3), consistent with a conclusion in \cite{niedzielski2010}. 
As may be seen in Table~1,  Fits~II and III  have large eccentricities. A
question remains, whether inferred orbital configurations are dynamically
stable. In fact, all these Fits~I--III, transformed to osculating elements at
the epoch of the first observation, lead to self-disruption of 2-planet
systems.   Nevertheless, remembering that stable solutions may still be found in
their neighborhood, we did dynamical analysis with the $N$-body, self-consistent
GAMP code \cite{gozdziewski2008} (which also relies on the GAs), trying to
refine Fits~I--III with the requirement of the long-term stability (the edge-on,
coplanar models are tested). Certainly, at most {\em one} of these models might
correspond to the real system. We did not found any stable orbits in the
vicinity of Fit~I. In a tiny neighborhood of Fit~II, there is a stable
configuration with $\Chi\sim 1.50$ and an rms $\sim 27$~m/s which, as the direct
numerical integrations show, is stable  over 1~Gyr.  The best result is found
for Fit~III as a stable solution, corresponding to the 2:1~MMR with $\Chi\sim
1.36$ and an rms of $\sim 25$~m/s, i.e., the same  as in the kinematic Fit~III.
The dynamical map \cite{gozdziewski2008} around this solution (Fig.~\ref{fig3},
{\em the right-hand panel}) reveals extended island of stability ($\sim
0.15$~au). This fit has moderate semi-amplitude librations ($\sim
15-30^{\circ}$) of the critical angles
$\theta_{1}=2\lambda_b-\lambda_c-\varpi_{b}$ (around $0^{\circ}$),
$\theta_{2}=2\lambda_b-\lambda_c-\varpi_{c}$ (around $180^{\circ}$), and
$\theta_3=\varpi_{b}-\varpi_{c}$ (around $180^{\circ}$).  The numerical
integrations confirmed that its stability is preserved at least over 1~Gyr. 
The 2:1~MMR Fit~III seems  the most promising planetary model explaining the RV
variability of the HD~240210. The 2:1~MMR is quite frequent in the sample of
$\sim 40$ known extrasolar systems with jovian planets, because 5--6
configurations were reported (see, http://exoplanet.eu). Hence, this new
system,  which could be the first one around evolved star, is likely. We stress
that solution~III is found in relatively extended stability zone, unlike Fit~II,
which lies in  a tiny, isolated area ($\sim 0.01$~au, Fig.\ref{fig3}, {\em the
left-hand panel}).  These two maps almost overlap in the $a_c$-range, hence
other, relatively extended stable islands are rather excluded in this region.
\begin{table}[h]
\caption{
Keplerian model parameters of 2-planet best fit solutions to the RV of
HD~240210. Formal measurement errors are rescaled by adding the stellar jitter 
of $\sigma_j=20$~m/s in quadrature. $T_0 \equiv 53,000$~days,
$N_v=N_{RV}-N_{p}=27$.
}
\smallskip
\label{3solution}
\centering
 \begin{tabular}{|c|rr|rr|rr|}
\hline
the best fit & \multicolumn{2}{c|}{I} 
	     & \multicolumn{2}{c|}{II} 
	     & \multicolumn{2}{c|}{III}\\
parameters   & {\bf b}   &  {\bf c} 
             & {\bf b}   &  {\bf c} 
	     & {\bf b}   &  {\bf c} \\
\hline
$P$~[day]        & 540 $\pm$ 29   & 441 $\pm$ 27 
                 & 484 $\pm$ 14   & 667 $\pm$ 54
                 & 485 $\pm$ 18   & 994 $\pm$ 97  \\
$K$~[m/s]        & 131 $\pm$ 29   &  85 $\pm$ 30 
                 & 147 $\pm$ 16   &  82 $\pm$ 36             
                 & 129 $\pm$ 30   & 63  $\pm$ 23 \\
$e$              &0.05 $\pm$ 0.09  &0.20 $\pm$ 0.14
                 &0.18 $\pm$ 0.14  & 0.74 $\pm$ 0.38              
                 &0.30 $\pm$ 0.13  & 0.58 $\pm$ 0.35\\
$\omega$ [deg]   & 204 $\pm$  51  & 356 $\pm$ 73 
                 & 261 $\pm$  41  & 36  $\pm$ 76                
                 & 302 $\pm$ 35   & 163 $\pm$ 87\\
$\tau-T_0$ [days]& 352 $\pm$ 65  & 610 $\pm$ 109
                 & 484 $\pm$ 45  & 502 $\pm$ 72               
                 & 536 $\pm$ 31  & 326 $\pm$ 156 \\
$V_{0} $[m/s]  & \multicolumn{2}{c|}{-3$\pm$7} 
& \multicolumn{2}{c|}{12 $\pm$ 9}  
&  \multicolumn{2}{c|}{22$\pm$ 12} \\
$\Chi$ & \multicolumn{2}{c|}{1.35} 
&  \multicolumn{2}{c|}{1.36}  
&  \multicolumn{2}{c|}{1.36} \\
rms [m/s]&  \multicolumn{2}{c|}{25.2} 
&\multicolumn{2}{c|}{25.2}  
& \multicolumn{2}{c|}{25.2}   \\
\hline
\end{tabular}
\end{table}

\section*{Conclusions}
Extrasolar planets hosted by  giant or evolved  stars bring important border
conditions for the planet formation theory.  In this work, we re-analysed the
literature RV data for evolved dwarf HD~240210, with our KFIT code relying on
quasi-global GAs. In the reasonable range of orbital periods less than
3600~days, we found three Keplerian solutions, which have the same $\Chi$ and an
rms.    By further dynamical analysis of these best-fit models, we selected the
most likely, stable solution, which corresponds to 2:1~MMR, and is located in
relatively extend zone of dynamical stability. Overall, if the 2-planet
configuration is assumed, the dynamical constraints seem rule out other two
models, but only new observations may confirm the 2:1~MMR hypothesis.

\medskip
\noindent {\bf Acknowledgements.}
This work is supported by Polish Ministry of Science, Grant
92/N-ASTROSIM/2008/0.

\end{document}